\documentclass[letter, 10pt, conference]{ieeeconf}
\IEEEoverridecommandlockouts
\usepackage{cite}
\usepackage{amsmath,amssymb,amsfonts}
\usepackage{algorithmic}
\usepackage{graphicx}
\usepackage{textcomp}
\usepackage{xcolor}
\usepackage[british]{babel}
\usepackage{hyperref}
\usepackage{fancyhdr}

\hypersetup{
    colorlinks=true,
    linkcolor=blue,
    filecolor=blue,      
    urlcolor=blue,
    citecolor=blue,
}
\usepackage{todonotes}
\usepackage{microtype}
\usepackage{subcaption}
\usepackage{booktabs} 
\usepackage{multirow}
\usepackage{mathtools}
\usepackage{orcidlink}

\title{\LARGE \bf
NetVAD: Foundation-Model Representation Learning for Identifier-Free Unsupervised Intrusion Detection
}

\author{Darren Fürst\,\orcidlink{0009-0006-3607-349X}$^{1}$, Patrick Levi\,\orcidlink{0000-0002-5216-4555}$^{1}$ and Sebastian Steindl\,\orcidlink{0000-0001-8141-5264}$^{1}$%
\thanks{$^{1}$All authors are with the Department of Electrical Engineering, Media and Computer Science, 
Ostbayerische Technische Hochschule Amberg-Weiden, Amberg, Germany. 
{\tt\small \{d.fuerst, p.levi, s.steindl\}@oth-aw.de}}%
}

\begin{document}

\fancypagestyle{firstpage}{
  \fancyhf{} 
  \renewcommand{\headrulewidth}{0pt} 
  \fancyfoot[L]{%
    \scriptsize 
    A version of this paper has been accepted for publication in the IEEE International Conference on Intelligence and Security Informatics (ISI) 2026. \\
    \copyright\ 2026 IEEE. Personal use of this material is permitted. Permission from IEEE must be obtained for all other uses, in any current or future media, including reprinting/republishing this material for advertising or promotional purposes, creating new collective works, for resale or redistribution to servers or lists, or reuse of any copyrighted component of this work in other works.%
  }
}

\maketitle
\thispagestyle{firstpage}
\pagestyle{empty}

\begin{abstract}
Detecting zero-day exploits in production networks requires robust Intrusion Detection Systems (IDS). However, current unsupervised models struggle to match the performance of supervised classifiers, which are trained for specific attacks only. To bridge this gap, we leverage the emerging capabilities of Network Foundation Models. We propose \textit{NetVAD}, a strictly identifier-free Variational Autoencoder that projects representations from a frozen Foundation Model into a task-specific latent space, trained solely on benign traffic. Evaluated on ToN-IoT and IoT-23, NetVAD achieves highly competitive performance with unsupervised training and supervised calibration. On ToN-IoT, it achieves a 98\,\% Micro F1-score and a 96\,\% Macro F1-score at an operational false positive rate. Unlike prior work, we show the model's performance transparently for all attack-classes of the datasets. While the architecture excels at discerning complex botnet behaviour (99.6\,\% F1 on Okiru), our evaluation reveals limitations of flow-based Foundation Models in detecting single-packet reconnaissance events. Finally, a comprehensive ablation study confirms that while large-scale pre-training is essential to prevent performance degrading, specialised decoder architectures are necessary to precisely model the complex benign manifold, ensuring attacks are caught more reliably, due to a higher reconstruction loss.\end{abstract}

\section{Introduction}
\label{sec:intro}
Despite rapid progress in Network Foundation Models (FMs), their utility for unsupervised intrusion detection remains largely unexplored. Existing FM-based network security research primarily focuses on supervised transfer learning~\cite{guthula2025netfoundfoundationmodelnetwork,lens, lin2022bert}, leaving open whether large-scale self-supervised traffic representations can enable identifier-free zero-day anomaly detection without attack labels or task-specific fine-tuning. To address this gap, we propose \textit{NetVAD}, a privacy-preserving framework that treats intrusion detection as a representation learning problem. Rather than learning directly from handcrafted flow statistics, NetVAD leverages frozen representations from the pre-trained \textit{netFound}~\cite{guthula2025netfoundfoundationmodelnetwork} transformer, which was trained on multi-modal network traffic using self-supervised objectives. These representations are projected into a task-specific variational latent space learned exclusively on benign traffic, enabling unsupervised anomaly detection without requiring exposure to attack samples during training. Consequently, NetVAD's capabilities can transfer to a zero-day detection setting, where attacks are identified as deviations from the learned model of benign network behaviour rather than through prior knowledge of specific attack signatures.

The approach operates under strictly identifier-free conditions, meaning we deliberately exclude IP-addresses, ports, timestamps, and payload content from the detector inputs. This prevents shortcut learning from network-specific identifiers while simultaneously preserving privacy and enabling deployment on encrypted traffic.

To evaluate the efficacy of our approach, we investigate the following Research Questions (RQs):
\begin{itemize}
    \item RQ1 (Efficacy): Can Foundation Models be used to successfully improve intrusion detection results in unsupervised, identifier-free settings?
    \item RQ2 (Necessity of Pre-training): To what extent does the large-scale pre-training of the Foundation Model contribute to the approach's success compared to training an identical Variational Autoencoder (VAE) from scratch?
    \item RQ3 (Adaptation): Which architectural choices are necessary to effectively exploit the complex learned representations of a Foundation Model for this unsupervised task?
\end{itemize}

In answering these questions, we make the following contributions:

\begin{itemize}
    \item We present one of the first studies, investigating whether pre-trained Network Foundation Models can enable fully unsupervised, identifier-free intrusion detection.
    \item We show several prior results for unsupervised learning approaches include features in their datasets, which may lead to shortcut learning, resulting in possibly inflated results. 
    \item We propose NetVAD, a FM adaptation architecture that combines frozen FM representations with a variational reconstruction objective to model the manifold of benign network behaviour.
    \item Through extensive ablations, we demonstrate that FM pre-training is essential for effective anomaly separation, while classic anomaly detectors applied directly to FM embeddings fail to exploit the learned representation space effectively.
    \item We show that NetVAD achieves highly competitive identifier-free IDS performance at operational false-positive rates while maintaining strong minority-class detection performance across heterogeneous attack families.
\end{itemize}

\section{Related Work}
\label{sec:related}
This section reviews the progression of unsupervised anomaly detection, the benchmarking challenges in intrusion detection, and the emergence of Network Foundation Models in security applications. Most modern intrusion detection systems are still primarily based on supervised classification models trained on labelled attack data. While effective in known attack scenarios, these approaches are inherently limited in their ability to generalise to unseen or zero-day attacks, motivating the need for unsupervised detection paradigms that learn normal behaviour without relying on attack labels, which we treat by designing NetVAD.

\begin{table*}[!ht]
\centering
\caption{Binary Performance Comparison of Unsupervised Models on ToN-IoT and IoT-23.}
\caption*{All Scores are Micro-Scores to be consistent with prior work. A '-' is placed for non-reported values. Best in \textbf{bold} and second best \underline{underlined}.}
\label{tab:baseline_comparison}
\resizebox{\linewidth}{!}{%
\begin{tabular}{lcccccccccc}
\toprule
& & & \multicolumn{6}{c}{\textbf{ToN-IoT}} & \multicolumn{2}{c}{\textbf{IoT-23}} \\
\cmidrule(lr){4-9} \cmidrule(lr){10-11}
\textbf{Model} & \textbf{Shortcut Risk?} & \textbf{Payload?} & \textbf{Precision} & \textbf{Recall} & \textbf{F1} & \textbf{FPR} & \textbf{ROC-AUC} & \textbf{PR-AUC} & \textbf{ROC-AUC} & \textbf{PR-AUC} \\
\midrule
Isolation Forest~\cite{zahoor2025robust} & \textcolor{red}{Yes} (IP/Port/Tstamp) & \textcolor{red}{Yes} & 0.37 & 0.08 & 0.13 & 63.06\% & - & - & - & - \\
OCSVM~\cite{zahoor2025robust} & \textcolor{red}{Yes} (IP/Port/Tstamp) & \textcolor{red}{Yes} & 0.84 & 0.99 & 0.91 & 16.04\% & - & - & - & - \\
\midrule
Isolation Forest~\cite{gorbett2022local} & No & \textcolor{red}{Yes} & - & - & - & - & 0.567 & 0.364 & 0.492 & 0.594 \\
KNN~\cite{gorbett2022local} & No & \textcolor{red}{Yes} & - & - & - & - & 0.973 & 0.943 & 0.970 & 0.970 \\
Autoencoder~\cite{gorbett2022local} & No& \textcolor{red}{Yes} & - & - & - & - & \underline{0.981} & 0.985 & \underline{0.981} & 0.972 \\
Weighted Hamming LID~\cite{gorbett2022local} & No & \textcolor{red}{Yes} & - & - & - & - & \textbf{0.985} & 0.983 & \textbf{0.998} & \textbf{0.994} \\
\midrule
Isolation Forest (Ours) & No & No & \underline{0.999} & \underline{0.905} & \underline{0.950} & \underline{5.94\%} & 0.953 & \underline{0.999} & 0.692 & \underline{0.991} \\
\textbf{NetVAD (Ours)} & No & No & \textbf{0.999} & \textbf{0.962} & \textbf{0.980} & \textbf{5.02\%} & 0.980 & \textbf{0.999} & 0.840 & 0.985 \\
\bottomrule
\end{tabular}%
}
\end{table*}

\subsection{Unsupervised IDS}\label{sec:unsuper_related}

Few works explicitly investigate unsupervised intrusion detection under identifier-free conditions that prevent shortcut learning from network-specific features such as IP-addresses, ports or timestamps. In many commonly used datasets, including ToN-IoT and CIC-IDS2017, attack traffic is generated by dedicated machines with static identifiers~\cite{TonIotDataset}. If such identifiers remain available during training, models may inadvertently exploit them instead of learning behavioural traffic characteristics. This phenomenon, commonly referred to as shortcut learning~\cite{geirhos2020shortcut}, can lead to overly optimistic results that do not reflect real-world generalisation, where an attacker’s traffic will not share the same identifiers as those seen during training. This is emblematic of a broader 'credibility crisis' in networking research, where few published ML models can be trusted to perform reliably in untested environments~\cite{willinger2025something}.
Graph-based approaches such as TGN-SVDD~\cite{liuliakov2023one} model network communication as dynamic temporal graphs in which endpoints are represented as nodes derived from IP-addresses and flows form timestamped edges. As noted by the authors, in the CIC-IDS2017 dataset attacker IP-addresses appear only during the testing phase, potentially enabling shortcut learning via node identity. To mitigate this issue, \cite{liuliakov2023one} inject additional events containing the attacker node into the benign training set. Additionally, because temporal graphs rely on timestamped event streams, models may capture dataset-specific temporal patterns, such as specific attack windows in the datasets, which may not generalise well to other networks, where attacks are not clearly bound to specific timeframes.
Verkerken et al.~\cite{verkerken2022towards}, explicitly investigate identifier-free intrusion detection by removing IP-addresses and timestamps prior to training. They evaluate several classic anomaly detection models. While their evaluation reports F1-scores and ROC-AUC curves, the false-positive rates corresponding to these scores are not explicitly provided, limiting assessment of performance at operational thresholds.

Several other studies evaluate unsupervised IDS models on modern benchmark datasets such as ToN-IoT and IoT-23. For example, \cite{zahoor2025robust} report that a One-Class SVM can achieve an F1-score of 0.91 on the ToN-IoT dataset. However, their subsequent feature importance analysis shows that the models rely heavily on network-specific identifiers, indicating a risk of shortcut learning. This makes it unclear how much of the performance stems from exploitation of such features. Similarly, \cite{gorbett2022local} evaluate multiple anomaly detection models, including Isolation Forests and Autoencoders, across both the ToN-IoT and IoT-23 datasets while proactively removing identifying features prior to training. This setup effectively mitigates shortcut learning and therefore provides a useful identifier-free baseline. However, their evaluation reports only aggregate metrics such as ROC-AUC and Precision-Recall AUC (PR-AUC), which do not directly allow comparison of model behaviour at operationally viable thresholds.

To address these limitations, NetVAD explicitly adopts a strictly identifier-free pipeline to prevent shortcut learning, and we evaluate our architecture at operationally viable FPR thresholds to provide a transparent, realistic assessment of deployment capability.

\subsection{Network Foundation Models}

For Network Foundation Models different approaches exist. Generative approaches such as \textit{NetGPT}~\cite{netgpt} and \textit{LENS}~\cite{lens} adapt GPT and T5 architectures for tasks ranging from traffic generation to few-shot classification. Other works target multi-modality and efficiency: \textit{MM4flow}~\cite{mm4flow} integrates traffic in two modalities: byte-streams and transmission patterns, whereas \textit{Traffic-MoE}~\cite{trafficmoe} uses a sparse Mixture-of-Experts architecture to decouple model capability from inference costs, to enable real-time traffic processing.

We adopt \textit{netFound}~\cite{guthula2025netfoundfoundationmodelnetwork}, a self-supervised hierarchical multi-modal transformer capturing traffic at the packet, burst, and flow levels, as our backbone encoder, due to their open-source availability and performant metrics for supervised transfer-learning, suggesting well-formed representations.

Existing Network Foundation Model research primarily evaluates representations in supervised settings, leaving their applicability to fully unsupervised, identifier-free anomaly detection largely unexplored. In particular, it remains unclear whether frozen self-supervised traffic embeddings can support effective modelling of normal network behaviour when only benign data is used for training.

\section{Methods}
\label{sec:methodology}

\subsection{NetVAD Architecture}

In this work, we treat the FM as a fixed representation backbone and learn a variational adaptation module that operates on the FM’s input token embedding space. Specifically, we extract the output of the FM's embedding layer, chosen to ensure tractable reconstruction by projecting variable-length multi-modal inputs into a structurally consistent, fixed-size representation and reconstruct these representations for benign traffic using a VAE approach. By optimising a reconstruction-based objective on benign samples only, the model learns a compact latent representation of the benign traffic in this embedding space. At inference time, deviations from this learned structure are identified through increased reconstruction error, enabling unsupervised anomaly detection without requiring exposure to attack data during training.

\subsubsection{Encoder Projection and Downsampling}
The netFound encoder outputs a sequence of representations. To compress this information into a task-specific latent space, we employ a learnable downsampling MLP. The downsampler consists of a sequence of Linear layers, each followed by Layer Normalisation~\cite{ba2016layer} and a ReLU activation. This structure progressively reduces the dimensionality into the target dimension of our latent space, forming vectors $\mu$ and $\sigma$, from which $z$ is sampled using the reparametrisation trick.

To preserve information lost during compression, which may be beneficial for reconstructing subtle benign traffic patterns with high fidelity, we implement U-Net~\cite{ronneberger2015u} style skip connections.

\subsubsection{Latent Distribution and Planar Flows}
To increase the expressiveness of the sampled posterior, we apply a sequence of Planar Flows~\cite{rezende2015variational} to the sampled latent vectors $\mathbf{z}$. Allowing the model to approximate more complex posterior distributions.
In practice, during hyperparameter tuning planar flows were not found to be beneficial to the model's performance.

\subsubsection{Decoder Head}
\begin{figure}[!ht]
    \centering
    \includegraphics[width=0.8\linewidth]{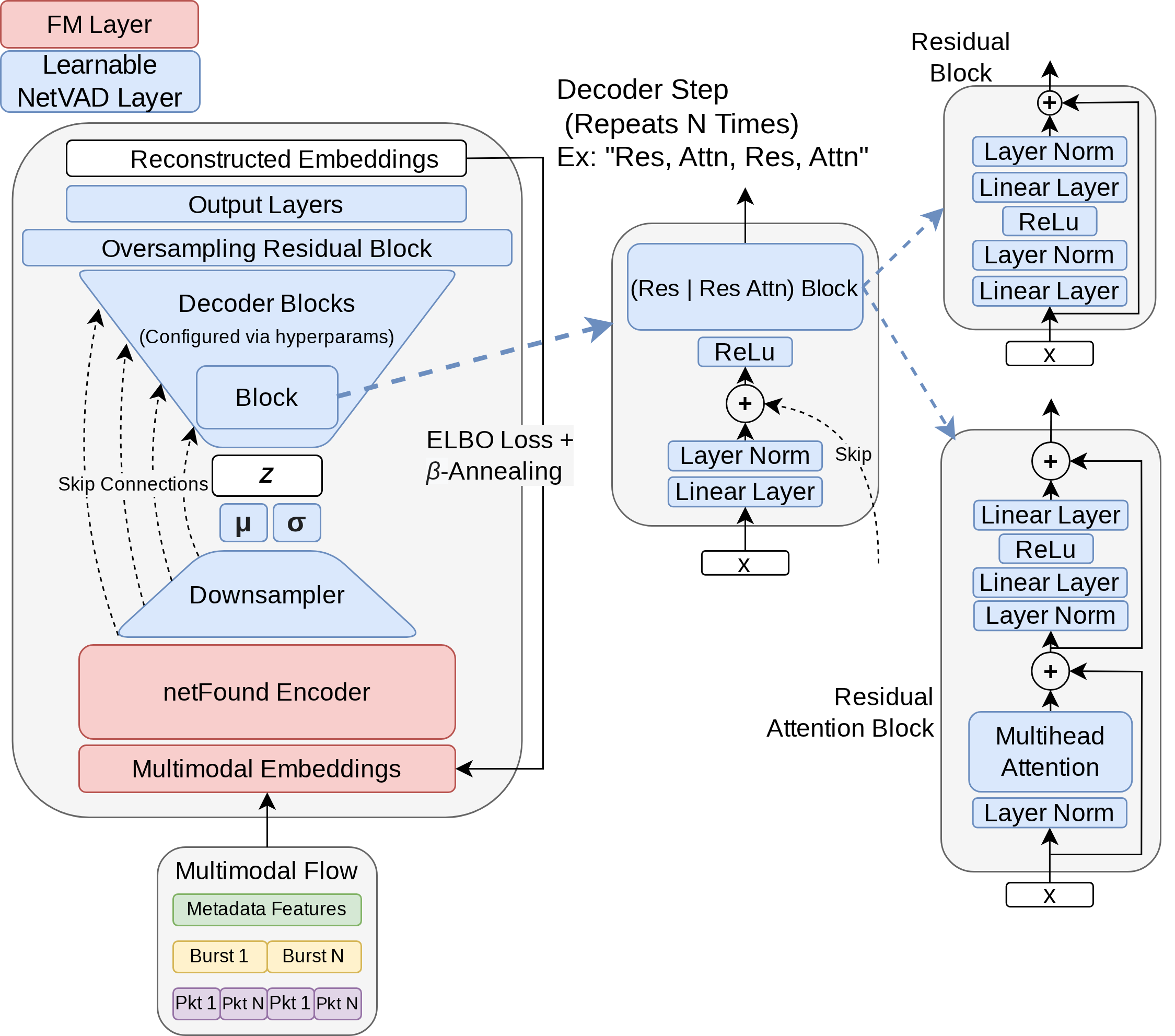}
    \caption{NetVAD Architecture}
    \label{fig:NetVAD_architecture}
\end{figure}

The decoder inverts the downsampling process, progressively upsampling the latent representation $z$ back to the size of the first embedding layer of the FM with dimensions $E \in \mathbb{R}^{L \times H}$. It is composed of linear upsampling layers followed by Layer Normalisation and ReLU activations. 
To refine the reconstruction, the upsampling path is interleaved with Residual Blocks. The Residual Block and Residual Attention Blocks are visualised in \autoref{fig:NetVAD_architecture}.
To enable the decoder to form richer representations, we oversample the reconstruction before the final output layer, inspired by feature map expansion techniques and wide architectures~\cite{sandler2018mobilenetv2, zagoruyko2016wide, schuster2025predtrad}. The oversampling amount is a tunable hyperparameter. 
The final output $\hat{E}$ aims to reconstruct the original input embeddings $E$ learned by the pre-trained netFound backbone.

\subsection{Loss Function}

To train NetVAD, we use the classic VAE ELBO loss function, consisting of a standard MSE-loss and a KL-divergence regularisation term, augmented with a $\beta$-annealing strategy introduced by \cite{CyclicalAnnealing}. This strategy cyclically anneals the KL-divergence term, enabling the model to construct a less regularised, meaningful latent space during the initial learning phase, before enforcing stricter regularisation.

\begin{equation} \label{eq:elbo}
    \mathcal{L} = \text{MSE}(\mathbf{E}, \mathbf{\hat{E}}) + \beta(t) \cdot \text{KL}(q(\mathbf{z}|\mathbf{E}) || p(\mathbf{z})).
\end{equation}

This is repeated in a cyclical fashion, allowing the model to repeatedly explore the latent space configuration during training, using each cycle's latent space as a ``warm restart''~\cite{CyclicalAnnealing}.

\section{Experimental Setup}

\subsection{Datasets and Pre-Processing}
To evaluate the robustness of our model across heterogeneous environments, we utilise raw network traffic (pcaps), from two widely used IoT security datasets: ToN-IoT~\cite{TonIotDataset} and IoT-23~\cite{garcia_2020_iot23}.

\subsubsection{ToN-IoT}
The ToN-IoT dataset includes network traffic from a diverse set of devices. We specifically utilise the raw network pcap recordings, encompassing nine distinct attack categories, such as DDoS, Ransomware, XSS, and Backdoors, launched against the infrastructure~\cite{TonIotDataset}.

\subsubsection{IoT-23}
The IoT-23 dataset provides $23$ pcaps representing a wide range of real-world IoT malware behaviours. It captures both traffic from botnets such as Mirai and Torii, executed on physical IoT hardware, and reconnaissance techniques such as portscans. As the label 'cncmirai' only has two flows in the dataset~\cite{garcia_2020_iot23}, we merge it into the 'cnc' class.

\subsubsection{Pre-Processing and Data Splits}
The raw pcap files were processed using the official netFound pre-processing pipeline. This extracts multi-modal features on the flow level, burst level, with a burst being multiple packets in a short time window and on the individual packet level. The burst and packet samples included for a network flow representation can be set as a hyperparameter. We follow~\cite{guthula2025netfoundfoundationmodelnetwork} in the selection of 12 representational bursts and packets. Importantly, the pre-processing discards the payload data to remain privacy-preserving, which allows inference on encrypted or unencrypted traffic alike. Specific identifiers such as IP-addresses, ports, and timestamps are excluded to prevent shortcut learning, ensuring the model learns generalisable robust traffic behaviour.

The dataset was partitioned into disjoint training, calibration, and test subsets. To transparently assess generalisation, the calibration set comprised exactly 1,287 benign and 2,000 attack flows, randomly sampled from the held-out distribution. The anomaly threshold was chosen to maximise the F1-score under a fixed false-positive-rate constraint (FPR $\leq$ 5\%) and was then frozen for evaluation on the final test set. This setup reflects an operational deployment scenario in which a detector is calibrated post-deployment using labelled validation traffic. For strictly unsupervised deployments lacking labelled data, thresholds could alternatively be derived from benign error percentiles or by leveraging synthetic attack generation.

\begin{table*}[!t]
\caption{Comprehensive Performance Metrics for NetVAD vs. Ablations marked with *. Table metrics are row-wise. Best F1 in \textbf{bold} and second best \underline{underlined}.}\label{tab:full_results}
\centering
\resizebox{\linewidth}{!}{%
\begin{tabular}{ll|cccc|cccc|cccc}
\toprule
 & & \multicolumn{4}{c}{\textbf{NetVAD}} & \multicolumn{4}{c}{\textbf{FM + Linear Layers only*}} & \multicolumn{4}{c}{\textbf{Isolation Forest* (Raw Features)}} \\ 
 \cmidrule(lr){3-6} \cmidrule(lr){7-10} \cmidrule(lr){11-14}
\textbf{Dataset} & \textbf{Attack Class} & \textbf{Prec.} & \textbf{Rec.} & \textbf{F1} & \textbf{AUC} & \textbf{Prec.} & \textbf{Rec.} & \textbf{F1} & \textbf{AUC} & \textbf{Prec.} & \textbf{Rec.} & \textbf{F1} & \textbf{AUC} \\ 
\midrule
\multirow{11}{*}{\rotatebox{90}{\textbf{ToN-IoT}}} 
 & \textbf{XSS} & 0.997 & 0.999 & \textbf{0.998} & 0.980 & 0.998 & 0.995 & \underline{0.997} & 0.978 & 0.997 & 0.720 & 0.836 & 0.914 \\
 & \textbf{Password} & 0.997 & 0.999 & \underline{\textbf{0.998}} & 0.990 & 0.998 & 0.998 & \underline{\textbf{0.998}} & 0.984 & 0.999 & 0.995 & 0.997 & 0.973 \\
 & \textbf{Backdoor} & 0.984 & 0.994 & \underline{0.989} & 0.980 & 0.987 & 0.994 & \textbf{0.990} & 0.972 & 0.981 & 0.993 & 0.987 & 0.952 \\
 & \textbf{Injection} & 0.997 & 0.998 & \textbf{0.998} & 0.980 & 0.998 & 0.996 & \underline{0.997} & 0.978 & 0.998 & 0.972 & 0.985 & 0.974 \\
 & \textbf{Scanning} & 0.996 & 0.999 & \underline{\textbf{0.998}} & 0.990 & 0.997 & 0.999 & \underline{\textbf{0.998}} & 0.997 & 0.995 & 0.985 & 0.990 & 0.942 \\
 & \textbf{Ransomware} & 0.904 & 0.993 & \underline{0.946} & 0.990 & 0.918 & 0.993 & \textbf{0.954} & 0.994 & 0.888 & 0.992 & 0.937 & 0.982 \\
 & \textbf{DDoS} & 0.997 & 0.965 & \underline{\textbf{0.981}} & 0.970 & 0.998 & 0.964 & \underline{\textbf{0.981}} & 0.967 & 0.998 & 0.880 & 0.935 & 0.949 \\
 & \textbf{DoS} & 0.996 & 0.800 & \textbf{0.887} & 0.960 & 0.996 & 0.641 & \underline{0.780} & 0.833 & 0.995 & 0.620 & 0.764 & 0.901 \\
 & \textbf{MITM} & 0.761 & 0.937 & \textbf{0.840} & 0.970 & 0.781 & 0.873 & \underline{0.824} & 0.965 & 0.605 & 0.535 & 0.568 & 0.862 \\ \cmidrule{2-14} 
 & \textit{\textbf{Macro}} & \textit{0.959} & \textit{0.965} & \textbf{\textit{0.959}} & \textit{0.980} & \textit{0.963} & \textit{0.939} & \underline{\textit{0.947}} & \textit{0.963} & \textit{0.940} & \textit{0.855} & \textit{0.889} & \textit{0.939} \\
 & \textbf{Micro} & \textit{0.999} & \textit{0.962} & \textbf{\textit{0.980}} & \textit{0.980} & \textit{0.999} & \textit{0.936} & \underline{\textit{0.966}} & \textit{0.957} & \textit{0.999} & \textit{0.905} & \textit{0.950} & \textit{0.953} \\ 
\midrule
\multirow{10}{*}{\rotatebox{90}{\textbf{IoT-23}}} 
 & \textbf{Okiru Botnet} & 0.997 & 0.995 & \textbf{0.996} & 0.990 & 0.991 & 0.866 & \underline{0.924} & 0.959 & 0.298 & 0.002 & 0.005 & 0.686 \\
 & \textbf{Generic Attack} & 0.965 & 0.938 & \underline{0.952} & 0.990 & 0.918 & 0.994 & \textbf{0.955} & 0.990 & 0.901 & 0.727 & 0.805 & 0.966 \\
 & \textbf{C\&C} & 0.958 & 0.624 & \textbf{0.756} & 0.880 & 0.900 & 0.645 & \underline{0.751} & 0.863 & 0.723 & 0.171 & 0.276 & 0.769 \\
 & \textbf{Port Scan} & 0.920 & 0.034 & \underline{0.066} & 0.680 & 0.854 & 0.045 & \textbf{0.085} & 0.691 & 0.947 & 0.017 & 0.033 & 0.689 \\
 & \textbf{C\&C + Scan} & 0.509 & 0.999 & \textbf{0.675} & 0.990 & 0.284 & 0.999 & \underline{0.442} & 0.995 & 0.078 & 0.194 & 0.111 & 0.942 \\
 & \textbf{File Download} & 0.146 & 0.999 & \textbf{0.255} & 0.990 & 0.061 & 0.999 & 0.116 & 0.998 & 0.065 & 0.980 & \underline{0.123} & 0.992 \\
 & \textbf{Torii Botnet} & 0.020 & 0.429 & \textbf{0.038} & 0.820 & 0.008 & 0.429 & 0.015 & 0.845 & 0.009 & 0.429 & \underline{0.017} & 0.811 \\
 & \textbf{C\&C + File} & 0.036 & 0.999 & \textbf{0.070} & 0.990 & 0.014 & 0.999 & 0.028 & 0.999 & 0.016 & 0.999 & \underline{0.031} & 0.999 \\ \cmidrule{2-14} 
 & \textit{\textbf{Macro}} & \textit{0.569} & \textit{0.753} & \textbf{\textit{0.476}} & \textit{0.920} & \textit{0.504} & \textit{0.747} & \underline{\textit{0.415}} & \textit{0.918} & \textit{0.380} & \textit{0.440} & \textit{0.175} & \textit{0.857} \\
 & \textbf{Micro} & \textit{0.998} & \textit{0.538} & \textbf{\textit{0.699}} & \textit{0.840} & \textit{0.993} & \textit{0.487} & \underline{\textit{0.653}} & \textit{0.834} & \textit{0.968} & \textit{0.023} & \textit{0.045} & \textit{0.692} \\
 \bottomrule
\end{tabular}
}
\end{table*}

\subsection{Metrics}
Many IDS datasets exhibit severe class imbalance, making Micro-scores prone to overestimating performance by being dominated by majority attack classes. For example, after pre-processing ToN-IoT contains 88,906 DoS flows but only 876 MITM flows. While we report Micro-scores to enable direct comparison with prior work, they can obscure poor performance on minority attack categories. Therefore, we additionally report Macro-scores and per-class metrics, providing a more balanced assessment of model behaviour across attack types. This is particularly important for evaluating generalisation, as strong aggregate performance may conceal weaknesses on specific attacks that could limit the detector's ability to identify unseen threats.

\subsection{Training and Hyperparameter Optimisation}
\label{sec:hyperparams}

Training was performed using the AdamW optimiser with a StepLR learning-rate scheduler. We employed cyclical $\beta$-annealing for the KL-divergence term~\cite{CyclicalAnnealing} and used early stopping with a patience of five epochs to prevent overfitting. Training was capped at a maximum of 60 epochs, although convergence occurred substantially earlier, after 23 epochs on ToN-IoT and 26 epochs on IoT-23.

To optimise the architecture for each dataset, we performed a hyperparameter search over latent dimensions $z \in \{16,\dots,512\}$, decoder blocks, and U-Net skip connections. Training dynamics, including learning rates, dropout probabilities, $\beta$-annealing cycles, and layer sizes, were jointly optimised. The search revealed distinct requirements for the two environments. The ToN-IoT dataset favoured a highly compressed latent space ($z=16$) with a standard sequential decoder, whereas IoT-23 required higher capacity ($z=256$) and U-Net skip connections. Although Planar Flows were included in the search space to increase posterior expressiveness, the best-performing configurations disabled this component. All hyperparameter optimisations were conducted using benign traffic only, with reconstruction error on normal traffic as the optimisation objective.

\section{Results and Discussion}
\label{sec:results}

We evaluate NetVAD on its ability to distinguish benign traffic from attacks without prior exposure to attacks. By treating network intrusion detection as a representation learning task, we demonstrate NetVAD's capacity to identify completely unseen threats through reconstruction errors on network data.

\subsection{Baseline Comparison}
To contextualise our approach, \autoref{tab:baseline_comparison} compares NetVAD against recent unsupervised anomaly detection models. Importantly, high-performing models, such as the OCSVM~\cite{zahoor2025robust} (Micro-F1: 0.91), achieve their results by retaining scenario- or network-specific identifiers like IP-addresses and ports, with its implications in regard to shortcut learning discussed in \autoref{sec:unsuper_related}. They also do not report their FPR when analysing their models, which we calculated from the provided metrics in their paper to be $16.04\,\%$. This further shows that our model is more suited for the task at a far lower FPR, which is necessary for IDS to mitigate alert fatigue by operators.

When compared to Gorbett et al.~\cite{gorbett2022local}, who trained their unsupervised models in an identifier-free manner similar to us, yet retaining payloads as a feature, NetVAD achieves highly competitive performance. While they report impressive aggregate metrics, such as a ROC-AUC of 0.985 on ToN-IoT and a PR-AUC of 0.994 on IoT-23, relying exclusively on aggregate scores obscures real-world operational viability. Aggregate ROC-AUC fail to reflect model performance at the strict, low FPR thresholds, while their PR-AUC is skewed by class imbalances. This is apparent on their reported per-class metrics, which are provided only for a subset of four classes on the more challenging IoT-23 dataset. For example, they report a PR-AUC of 0.989 for Portscans, compared to NetVAD's 0.894. Conversely, for the Okiru Botnet class, NetVAD achieves a substantially higher PR-AUC of 0.991, whereas their reported PR-AUC is 0.701. A similar pattern emerges on ToN-IoT. While their per-class PR-AUC scores fall below 0.9 for all but one attack class, NetVAD achieves a PR-AUC above 0.98 for every class except MITM.
These disparities in aggregated metrics show once more, that for effective evaluation of such models metrics for each class at a transparent FPR must be published. Finally, NetVAD isolates these advanced threats in a strictly privacy-preserving manner without inspecting payload data, whereas the baseline's reliance on payloads as features fundamentally limits its deployability on modern encrypted networks.

It is also worth noting that our own implementation of an identifier-free Isolation Forest (IF) baseline (detailed in our ablation study, \autoref{tab:full_results}) outperforms the IF benchmark reported by Gorbett et al.~\cite{gorbett2022local} (e.g., achieving a Micro ROC-AUC of 0.953 vs. 0.567 on ToN-IoT and 0.692 vs. 0.492 on IoT-23) and the IF benchmark by Zahoor et al.~\cite{zahoor2025robust} (e.g., Micro-F1 of 0.95 vs. 0.13). We attribute these substantial improvements entirely to the multi-modal flow preprocessing provided by the netFound~\cite{guthula2025netfoundfoundationmodelnetwork} ecosystem.
By capturing traffic dependencies across packet, burst, and flow levels, this tokenisation strategy extracts a richer, more robust statistical representation, even before passing the data through the FM's transformer blocks.

\subsection{ToN-IoT Evaluation}
As detailed in \autoref{tab:full_results}, our NetVAD achieves an overall Macro F1-score of 95.9\% and a Recall of 96.5\% on the ToN-IoT dataset. The model demonstrates high robustness against different attacks, achieving over 99.8\% F1-scores for XSS, Password, and Backdoor attacks. NetVAD detects these to a high degree, as the model achieves significantly higher reconstruction errors for these attacks compared to benign data, as the attacks fail to map onto the tightly constrained benign manifold in the latent space. Conversely, performance on DoS and MITM attacks is comparatively lower (F1: 88.7\% and 84.0\%). DoS and MITM attacks often exhibit low interaction complexity, with limited packet exchanges and highly repetitive structures. As a result, their representations in the frozen FM embedding space may exhibit low variance and can overlap with benign traffic patterns. Since the model reconstructs flow-level representations independently and is primarily optimised on benign traffic, such low-complexity attack flows may not induce sufficiently distinct deviations, leading to comparatively lower reconstruction error. This suggests that reconstruction-based detection in the FM embedding space is more sensitive to structurally diverse attacks than to low-entropy, highly regular traffic patterns.
Recommendations based on this insight are given in \autoref{sec:conclusion}.

\subsection{IoT-23 Evaluation}
On the IoT-23 dataset, the global metrics reflect the challenging nature of the traffic. While the Micro F1-score is 69.9\%, the Macro F1-score drops to 47.6\%. However, for specific advanced threats, the model remains highly effective. Most notably, NetVAD achieves an F1-score of 99.6\% on Okiru botnet traffic, proving the architecture can isolate complex, botnet behaviour with high precision. In contrast, Port Scans exhibit a very poor F1-score (6.6\%). The scans in the IoT-23 dataset are SYN-scans, which are single-packet events. Because these flows are structurally simple, NetVAD easily reconstructs them with a low error margin, causing them to be misclassified as benign traffic. We discuss potential mitigations for this limitation in~\autoref{sec:conclusion}.
The scans in the ToN-IoT dataset were performed with noisier ``nmap'' and ``Nessus'' scans~\cite{TonIotDataset}, which explains why these can be detected more easily.
Similarly, Torii Botnet (F1: 3.8\%) and File Download (F1: 25.5\%) evade detection. While we suspect their representations heavily overlap with benign traffic, a detailed qualitative structural analysis of these specific failure modes remains an area for future investigation.

To explicitly answer RQ1, these combined results demonstrate that FMs can indeed successfully enable unsupervised, identifier-free intrusion detection. While limitations exist for single-packet events, our highly competitive performance on ToN-IoT (98.0\,\% Micro-F1) and our ability to precisely isolate sophisticated botnets on IoT-23 validate that frozen FM representations provide a robust feature space for zero-day anomaly detection without relying on dataset-specific shortcuts.

\subsection{Ablation Study}
To validate the necessity of the Foundation Model and to evaluate our custom VAE architectural choices, we conducted a multi-part ablation study, detailed in \autoref{tab:full_results}.

To answer RQ2, we trained the full NetVAD architecture from scratch with random weight initialisation, completely bypassing the pre-trained FM. This resulted in a drastic performance drop. The Macro F1-score went from 95.9\,\% to 19.0\,\% on ToN-IoT, and from 47.6\,\% to 16.7\,\% on IoT-23. This answers RQ2, showing that large-scale pre-training is essential for constructing a meaningful latent space capable of separating anomalies from normal traffic in such an unsupervised setting.

To answer RQ3, we replaced our specialised VAE decoder (which utilises U-Net skip connections, residual attention blocks, and oversampling) with linear layers only for both the downsampler added to further compress the FM embeddings and the decoder (\textit{FM + Linear Layers only}). On the ToN-IoT dataset, this basic VAE retained a competitive Macro F1 of 94.7\,\%. This demonstrates that the Foundation Model successfully clusters general behavioural representations in the latent space, requiring minimal architectural complexity to decode on simpler datasets. However, on the significantly more structurally complex IoT-23 dataset, the limitations of a basic linear architecture became apparent. Specific complex classes saw drastic drops in performance. For example, detection of C\&C + Scan attacks dropped from 67.5\,\% to 44.2\,\%, and the Okiru Botnet dropped from 99.6\,\% to 92.4\,\%. This confirms that while the Foundation Model provides a robust feature baseline, our specialised architectural strategies are necessary to precisely model the benign manifold. By reconstructing normal traffic with higher fidelity, the advanced decoder ensures that a greater number of attacks fail to reconstruct, thereby amplifying their anomaly scores and more effectively separating them from benign behaviour. This answers RQ3 and shows that a more complex model architecture is necessary to improve the model's capability to distinguish attacks from regular traffic.

\subsubsection{NetVAD vs. Classic Approaches}
Finally, to determine whether a classic anomaly detector could substitute the VAE, we evaluated an IF using two distinct input representations: the raw multi-modal flow features and the frozen FM embeddings. As shown in \autoref{tab:full_results}, the IF trained on raw features achieved a Macro F1-score of 88.9\,\% on ToN-IoT and 17.5\,\% on IoT-23. In our experiments applying the IF directly to the FM embeddings degraded performance significantly, scoring an overall Macro F1 of only 54.0\,\% on ToN-IoT and 14.0\,\% on IoT-23, which is why the raw ablation is shown in the table. The IF failing to successfully exploit the FM representations to accomplish the task, suggests that the downsampling layers added to NetVAD, to allow the model to restructure the FM embeddings to the task, may aid in our model's performance. Furthermore, it is noteworthy that the IF trained solely on the raw features outperforms several existing unsupervised models reported in the literature (as compared in \autoref{tab:baseline_comparison}). We attribute this to the multi-modal flow pre-processing pipeline, which extracts a highly robust and discriminative feature set in its own right, providing a strong anomaly detection baseline even before the application of deep representation learning.

\section{Conclusion}\label{sec:conclusion}

In this paper, we presented NetVAD, a novel approach to unsupervised network intrusion detection by coupling a FM with a custom VAE architecture. We achieve competitive results to prior works on our tested datasets and provide thorough ablation results for the architecture and method. Importantly, the approach operates under strictly identifier-free conditions, preventing shortcut learning from network-specific features such as IP addresses or ports. Future work will focus on addressing the identified limitations regarding single-packet and low-volume events. Specifically, we intend to move beyond bidirectional flows, formed by source and destination tuples toward aggregate views, providing the model with the necessary behavioural context to identify stealthy scans and distributed attacks. Furthermore, we will explore and compare the viability of various FMs as pre-trained encoders across a broader range of network environments and datasets.

\section{Limitations}
This work evaluates the novel approach of constructing a VAE from pre-trained network FMs solely using a single FM on two widely used IDS datasets. This limits assessment of the approach's viability with different FMs, as well as on further network datasets. During testing, a blind-spot in regard to single or low-packet events was found. As discussed in the Conclusion, this limitation needs to be addressed in the future by providing the model with auxiliary information surrounding the flow of interest. Deploying this model in production requires evaluating inference latency, model size, and potential quantisation or knowledge distillation to reduce the combined model size of the architecture. These operational necessities were not explored in this work.


\section{Generative AI Use Disclosure}
Generative AI was used for editing and grammar correction. We are responsible for the content and quality of this paper.


\bibliographystyle{IEEEtran}
\bibliography{literature}

\section*{OCSVM Evaluation and Scalability}

To further contextualise our baseline selection, we conducted secondary experiments using a One-Class Support Vector Machine (OCSVM). Because OCSVMs scale quadratically ($\mathcal{O}(n^2)$) with the number of samples, training on the complete set of normal flows was computationally intractable. Consequently, the OCSVM was trained on a randomly sampled subset of 25,000 normal flows.

When evaluated on raw, identifier-free flow statistics, the OCSVM yielded poor detection results, achieving a Macro F1-score of 45.8\,\% on ToN-IoT and 0.18\,\% on IoT-23. This indicates that without network shortcut identifiers (e.g., IP-addresses), the overlapping distributions of raw flow statistics hinder the ability to establish a clear decision boundary. Interestingly, when the OCSVM was supplied with the Foundation Model embeddings, performance improved to a 61.2\,\% Macro F1 on ToN-IoT and 33.9\,\% on IoT-23. This supports the notion that the Foundation Model effectively clusters behavioural representations in the latent space. However, because the overall performance remained substantially lower than both the Isolation Forest and NetVAD, the Isolation Forest was used as a primary baseline for our study.

\section*{Computational Footprint}
To provide transparency regarding the operational costs of our approach, we report the parameter counts for our two configurations. The NetVAD models comprise approximately 1.29 billion parameters for ToN-IoT and 0.81 billion for IoT-23. In both instances, the pre-trained Foundation Model backbone accounts for $\approx 781$ million parameters, while the remaining capacity resides in the custom variational decoder and downsampler. It is important to note that the variation in parameter counts is a result of extensive hyperparameter tuning; the compact decoder size for IoT-23 (29 million parameters) was not an architectural constraint, but the optimal configuration identified to best minimise reconstruction error on the benign manifold during tuning. It is of course notable, that the ToN-IoT results with the bigger model are vastly superior to the IoT-23 performance. However, as the tuning objective was optimal reconstruction of benign traffic, it is not possible to tell whether a larger model would have reconstructed the anomalies in a more disjunct manner than the smaller model. Future work could explore this, by deliberately cross-comparing found configurations between datasets. 


\end{document}